\documentclass[twocolumn,aip,reprint]{revtex4-1}
\usepackage[latin9]{inputenc}
\usepackage[a4paper]{geometry}
\usepackage{xcolor}
\usepackage{float}
\usepackage{amsmath}
\usepackage{amssymb}
\usepackage{graphicx}

\makeatletter

\DeclareFontEncoding{LGR}{}{}
\DeclareRobustCommand{\greektext}{%
  \fontencoding{LGR}\selectfont\def\encodingdefault{LGR}}
\DeclareRobustCommand{\textgreek}[1]{\leavevmode{\greektext #1}}
\ProvideTextCommand{\~}{LGR}[1]{\char126#1}

\usepackage{dcolumn}
\usepackage{bm}
\usepackage{epsfig}
\usepackage{braket}

\bibpunct{(}{)}{,}{n}{}{}

\setcitestyle{round}
\renewcommand{\fnum@figure}{\textbf{Fig.~\thefigure}}

\makeatother

\begin{document}
\title{Echo in a Single Molecule}
\author{Junjie Qiang}
\thanks{JQ and IT contributed equally to this work.}
\affiliation{State Key Laboratory of Precision Spectroscopy, East China Normal
University, Shanghai 200062, China}
\author{Ilia Tutunnikov}
\thanks{JQ and IT contributed equally to this work.}
\affiliation{AMOS and Department of Chemical and Biological Physics, Weizmann Institute
of Science, Rehovot 7610001, Israel}
\author{Peifen Lu}
\affiliation{State Key Laboratory of Precision Spectroscopy, East China Normal
University, Shanghai 200062, China}
\author{Kang Lin}
\affiliation{State Key Laboratory of Precision Spectroscopy, East China Normal
University, Shanghai 200062, China}
\author{Wenbin Zhang}
\affiliation{State Key Laboratory of Precision Spectroscopy, East China Normal
University, Shanghai 200062, China}
\author{Fenghao Sun}
\affiliation{State Key Laboratory of Precision Spectroscopy, East China Normal
University, Shanghai 200062, China}
\author{Yaron Silberberg}
\affiliation{AMOS and Department of Physics of Complex Systems, Weizmann Institute
of Science, Rehovot 7610001, Israel}
\author{Yehiam Prior}
\email{yehiam.prior@weizmann.ac.il}

\affiliation{AMOS and Department of Chemical and Biological Physics, Weizmann Institute
of Science, Rehovot 7610001, Israel}
\affiliation{State Key Laboratory of Precision Spectroscopy, East China Normal
University, Shanghai 200062, China}
\author{Ilya Sh. Averbukh}
\email{ilya.averbukh@weizmann.ac.il}

\affiliation{AMOS and Department of Chemical and Biological Physics, Weizmann Institute
of Science, Rehovot 7610001, Israel}
\author{Jian Wu}
\email{jwu@phy.ecnu.edu.cn}

\affiliation{State Key Laboratory of Precision Spectroscopy, East China Normal
University, Shanghai 200062, China}
\affiliation{Collaborative Innovation Center of Extreme Optics, Shanxi University,
Taiyuan, Shanxi 030006, China}
\begin{abstract}
Echo is a ubiquitous phenomenon found in many physical systems, ranging
from spins in magnetic fields to particle beams in hadron accelerators.
It is typically observed in inhomogeneously broadened ensembles of
nonlinear objects, and is used to eliminate the effects of environmental-induced
dephasing, enabling observation of proper, inherent object properties.
Here, we report experimental observation of quantum wave packet echoes
in a \textit{single isolated molecule}. In contrast to conventional
echoes, here the entire dephasing-rephasing cycle occurs within a
single molecule without any inhomogeneous spread of molecular properties,
or any interaction with the environment. In our experiments, we use
a short laser pulse to impulsively excite a vibrational wave packet
in an anharmonic molecular potential, and observe its oscillations
and eventual dispersion with time. A second delayed pulsed excitation
is applied, giving rise to an echo - a partial recovery of the initial
coherent wavepacket. The vibrational dynamics of single molecules
is visualized by time-delayed probe pulse dissociating them one at
a time. Two mechanisms for the echo formation are discussed - ac Stark-induced
molecular potential shaking and creation of depletion-induced ``hole''
in the nuclear spatial distribution. Interplay between the optically
induced echoes and quantum revivals of the vibrational wave packets
is observed and theoretically analyzed. The single molecule wave packet
echoes may lead to the development of new tools for probing ultrafast
intramolecular processes in various molecules.
\end{abstract}
\maketitle
Echoes in physical systems, from their inception by Erwin Hahn in
1950 \cite{Hahn1950,Hahn1953}, were born within the framework of
spins embedded in an inhomogeneous environment, and are typically
used to measure dephasing rates due to the local interaction of the
spins (molecules) with their surroundings. Following the original
discovery of echoes in ensembles of spins excited by pulsed magnetic
fields, many different types of echoes have been observed, including
photon echoes \cite{Kurnit1964,Mukamel1995} and their mechanical
analogs \cite{Chebotayev1983}, cyclotron echoes \cite{Hill1965},
plasma-wave echoes \cite{Gould1967}, cold atoms echoes in optical
traps \cite{Bulatov1998,Ertmer2000,Herrera2012}, echoes in cavity
QED \cite{Meunier2005}, echoes in particle accelerators \cite{Stupakov1992,Spentzouris1996,Stupakov2013,Sen2018},
and more recently - echoes in a gas of rotating molecules \cite{Karras2015-2,Karras2016,Lin2016,Fleischer2016,Fleischer2017,Fleischer2018}.
In general, an ensemble of nonlinear objects (spins, molecules, plasma,
beam of particles, etc.) is impulsively excited by an external stimulus
resulting in a prompt coherent response. Since each member of the
ensemble evolves at a different frequency, the molecular response
disappears as the members of the ensemble step out of phase. However,
and this is the essence of the echo phenomenon, the effect of the
first excitation is not lost, and it is possible to retrieve it. By
applying a second pulsed excitation and waiting a time period equal
to the delay between the two excitations, a response signal emerges
again, and is referred to as an ``echo''. In general (but not in
our case, see below), interaction with the environment causes loss
of coherence, and reduces the echo signal amplitude. This reduction
in the echo amplitude provides information on the environmentally-induced
decoherence. The echo in an ensemble of nonlinear objects should be
distinguished from quantum revivals. In contrast to echoes, the quantum
revivals appear after a single-pulse excitation, and their period
depends only on the intrinsic properties of the molecular energy spectrum
\cite{Eberly1980,Parker1986,Averbukh1989,Robinett2004}.

Distinctly and in parallel, the interest in single molecule spectroscopy
has increased starting with the seminal works of W. E. Moerner \cite{Moerner1989}
and Michel Orrit \cite{Orrit1990}, and eventually leading to super-resolution
and super-sensitive optical microscopy. Single molecules embedded
in thin films are selected spectrally by narrow band CW lasers within
homogeneously broadened lines, or isolated spatially by observing
dilute samples where only one molecule lies within the field of view
of the microscope. An alternative approach to observing single molecules
is to measure them embedded within individual quantum dots, and then
spectral or spatial resolutions are not required, enabling time domain
experiments \cite{Lienau2002,Lienau2004}. Ultrafast experiments based
on sensitive fluorescence detection allowed observation of various
coherent excitations of vibrational wave packets in single molecules
\cite{Brinks2010} (for a review see \cite{hulstreview2014} and a
more recent work \cite{hulst2018} and references therein). Coherent
oscillatory signals from molecular vibrational wave packets impulsively
excited by short pulses are typically washed out on the femtosecond
time scale after several oscillations. Such a decay is not caused
by the ensemble inhomogeneity, and it has nothing to do with interaction
with the environment on such a short time scale. The collapse of the
coherent transients originates from intramolecular dephasing of multiple
vibrational states forming the wave packet in a single molecule. This
is, actually, a quantum phenomenon, as oscillations of a single nonlinear
classical oscillator never collapse, and last ``forever''.

The other type of ``single particle experiment'' is typically an
interference experiment where a single electron, or a single photon,
passes through a double slit and interference is observed due to the
interaction of this single particle with the double slit (\cite{Bach2013,Aspect2019},
and references therein). Each particle passes the double slit and
reaches the screen behind it without any other particles present,
and the probability of its hitting the screen at a specific location
is provided by quantum mechanics. The probability distribution can
be measured by repeating the measurement many times, every time with
a new single particle. These experiments are known as ``single photon''
or ``single electron'' interference experiments, and this is exactly
the type of experiment we are presenting here, but in our case the
interference happens in the time domain. The molecules interact with
the laser fields one at a time and are individually measured. As in
similar experiments, in order to see the distributions, the measurement
is repeated many times.

In the present paper, we show that it is possible to partially overcome
the effects of intramolecular dephasing and recur the signal by inducing
an echo in a single isolated molecule. We report experimental observation
of the quantum Wave Packet Echo (WPE) at the level of a single molecule,
and provide a detailed theoretical analysis of this phenomenon. In
our experiments, using an ultrashort femtosecond laser pulse, we create
a localized vibrational wave packet in an argon dimer cation. This
wave packet oscillates back and forth in the ion potential well and
disperses completely due to the spread of the oscillation frequencies
within the wave packet. Next, after a time delay of $t_{k}$, a second
laser pulse is applied. We demonstrate that an impulsive molecular
response (echo) indeed emerges at time $2t_{k}$ as a result of the
rephasing of the components of the wave packet. The entire sequence
of excitation, dispersion, re-excitation and the WPE occurs within
a single isolated molecule. The ability to observe a single molecule
event stems from our detection methodology of COLTRIMS (COLd Target
Recoil Ion Momentum Spectroscopy) \cite{Dorner2000,Ullrich2003} (see
below and in the Methods section), where the dissociation fragments
of each single molecule are individually detected following its dissociation.
The observed WPE is a generic phenomenon that can, in principle, be
observed by imaging nuclear dynamics in a variety of molecules \cite{De2011,Litvinyuk2011},
and in other impulsively excited isolated quantum systems. Our choice
of weakly bound argon dimers is motivated by their relatively low
vibrational frequencies which simplifies the time-resolved measurements.\\

\begin{figure}[h]
\begin{centering}
\includegraphics[width=0.9\columnwidth]{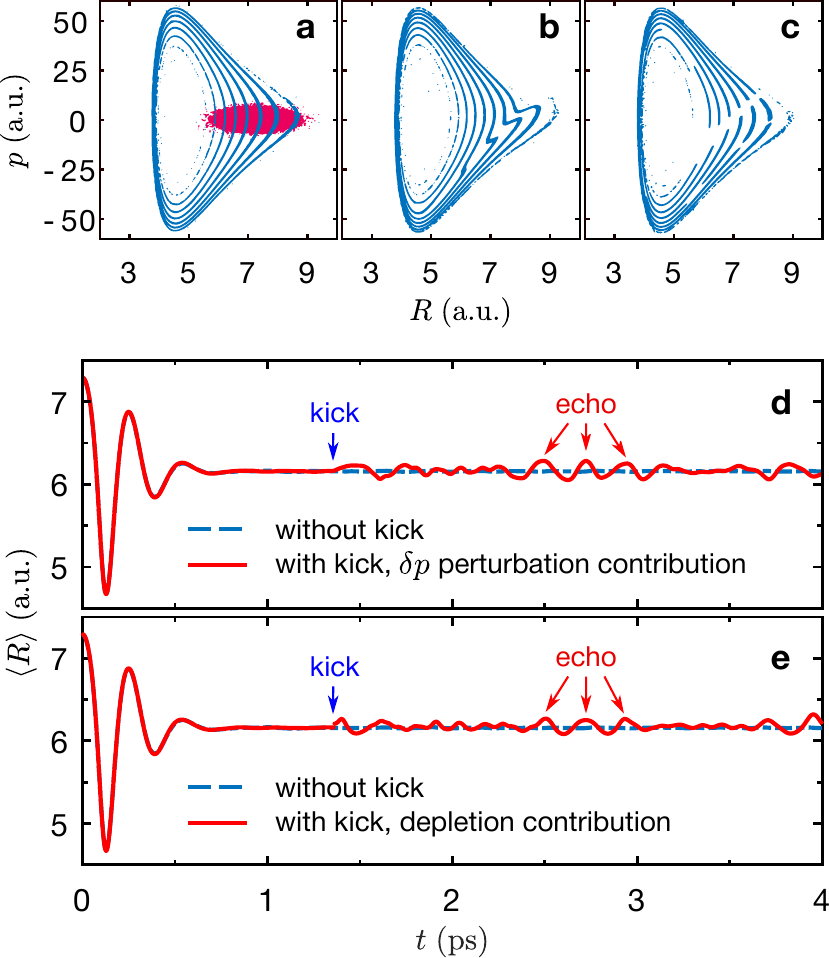}
\par\end{centering}
\caption{\textbf{Phase space dynamics. }The temporal evolution of the non-equilibrium
phase space distribution (Eq. \ref{eq:initial-distribution-1-1})
of an ensemble of molecules ($N=10^{5}$ ). \textbf{a}, Initial distribution
(in light red) and filamented phase space distribution just before
the second kick at $t_{k}=1.36$ ps (in blue), \textbf{b}, The distribution
shortly after a kick consisting of a uniform momentum increment in
each molecule by $\delta p=1$ a.u., and \textbf{c}, The distribution
shortly after a kick consisting of depleting molecules within the
strip of 5.5 a.u. \textless{} R \textless{} 6.1 a.u.. The next two
panels illustrate the ensemble averaged interatomic separation, $\braket{R}$$\left(t\right)$
and the echo response at $t\approx2t_{k}$ without (dashed blue) and
with (solid red) a kick: \textbf{d}, depicts the contribution of the
$\delta p$ perturbation (corresponding to the filamented phase space
in panel b) and \textbf{e}, depicts the contribution of the particles
depletion (corresponding to the filamented phase space in panel c).
\label{fig:Fig1}}
\end{figure}

\begin{figure}[h]
\begin{centering}
\includegraphics[width=0.9\columnwidth]{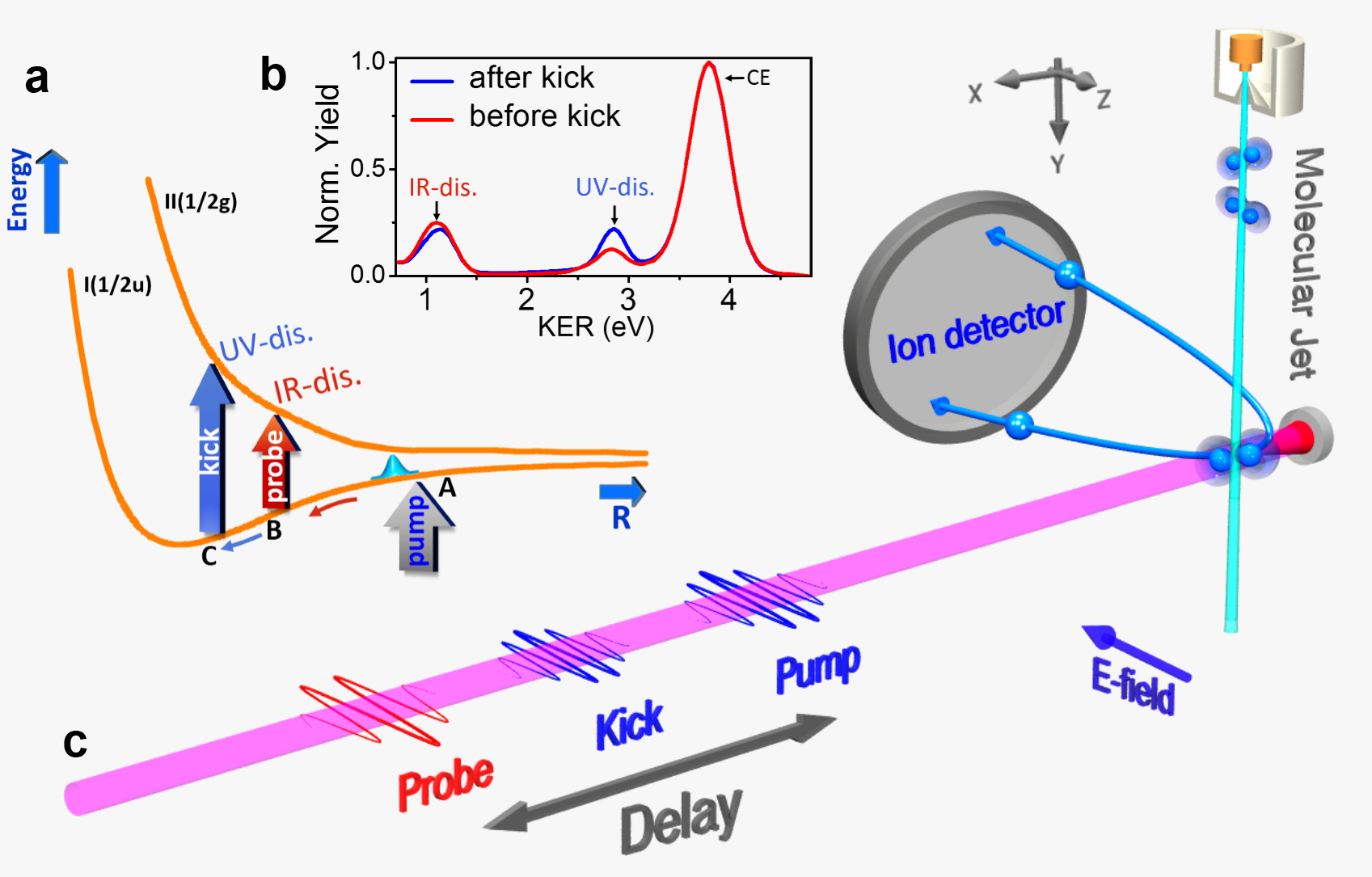}
\par\end{centering}
\caption{\textbf{Experimental setup.} \textbf{a}, A coherent nuclear wave packet
is launched using the pump pulse which excites a neutral argon dimer
$\mathrm{Ar}_{2}$ (not shown) to the $\mathrm{I}\left(1/2u\right)$
potential of the ion $\mathrm{Ar}_{2}^{+}$ at point A. The wave packet
oscillates in the bound $\mathrm{I}\left(1/2u\right)$ potential,
and is later dissociated by the excitation of the ion to its dissociating
$\mathrm{II}\left(1/2g\right)$ potential curve. The excitation to
the $\mathrm{II}\left(1/2g\right)$ state can be done by either the
probe pulse at 790 nm at point B, or by the kick pulse at 395 nm at
point C, leading to dissociation via the $\mathrm{Ar}_{2}\left(1,0\right)$
channel with different KER ranging from 0.7 eV to 3.2 eV as shown
in \textbf{b}, In the above, ``IR-dis.'' and ``UV-dis.'' refer
to dissociation induced by 790 nm and 395 nm pulses, respectively.
At higher energies, close to $4\;\mathrm{eV}$, the Coulomb-explosion
(CE) double ionization peak is seen. \textbf{c}, Schematic illustration
of the experimental apparatus. \label{fig:Fig2}}
\end{figure}

\noindent \textbf{\textcolor{teal}{Classical counterpart of the quantum
WPE.}} While the vibrational WPE in a single molecule is a quantum
phenomenon, a classical phase space analysis of the evolution of an
ensemble of molecules provides useful insights into the mechanism
of the echo formation. With this qualifying remark in mind, we now
consider the vibrational dynamics in an ensemble of non-interacting
homonuclear diatomic molecules. The vibrational motion of each molecule
is modeled as motion of a classical point particle having mass $M/2$
in a one-dimensional anharmonic potential, where $M$ is the mass
of each atom. Motivated by our experiments, we consider the $\mathrm{I}\left(1/2u\right)$
potential of argon dimer cation $\mathrm{Ar}_{2}^{+}$ (Fig. \ref{fig:Fig2}a).
The potential curve is approximated by a Morse potential, $V\left(R\right)=D\left(1-\exp\left[-a\left(R-R_{0}^{+}\right)\right]\right)^{2}$,
where $R$ is the interatomic separation, and the parameters are $D=0.048$,
$a=0.85$, and $R_{0}^{+}=4.56$ (all in atomic units) \footnote{The three parameters are obtained from fitting the \emph{ab initio
}potential curve \cite{Ar2PlusPotential} with Morse potential.}. The Hamiltonian governing the motion of each molecule is $\mathcal{H}\left(p,R\right)=p^{2}/M+V\left(R\right)$,
where $p$ is the momentum associated with the vibrational motion.
To imitate the initial state created by an instantaneous ionization
of $\mathrm{Ar}_{2}$, the initial distribution is described by a
non-equilibrium phase space distribution (Fig. \ref{fig:Fig1}a, light
red)
\begin{equation}
\rho\propto\exp\left[-\left(R-R_{0}\right)^{2}/2\sigma^{2}\right]\exp\left[-\sigma^{2}p^{2}/2\right],\label{eq:initial-distribution-1-1}
\end{equation}
where $R_{0}$ is the equilibrium internuclear separation of $\mathrm{Ar}_{2}$,
and $\sigma^{2}$ is the ground state variance of $\mathrm{Ar}_{2}$
(obtained by Gaussian function fit).

In an anharmonic potential, the frequency (period) of oscillations
is energy-dependent. As a result, the initial smooth phase-space distribution
(Fig. \ref{fig:Fig1}a, light red) evolves with time into a spiral-like
structure (blue) and gradually fills the accessible phase space as
shown in Fig. \ref{fig:Fig1}a. This filamentation of the phase space
results in an increasing number of spiral turns which become thinner
to conserve the volume, as is known in stellar systems \cite{Lynden-Bell1967}
and in accelerator physics \cite{Lichtenberg1969}. To induce the
echo response, a second excitation (``kick'') is applied at $t=t_{k}$
when the filamentation is well-developed. In our experiments, the
kick is provided by an ultrashort laser pulse that disturbs the molecules
in two ways: (i) it instantaneously changes the nuclei momentum due
to the coordinate-dependent ac Stark shift of the molecular potential,
and (ii) it transfers a fraction of the ground state molecules to
the excited electronic state via the optical absorption process. In
the latter case, the kick drills a ``hole'' in the spatial distribution
(a vertical strip in the phase space distribution), centered at the
resonance position \cite{Banin1994}. We consider the two mechanisms
separately and show that both lead to a subsequent echo response at
$t=2t_{k}$ as shown in Figs. \ref{fig:Fig1}(d,e).

In the case of the first mechanism, a kick caused by coordinate-dependent
ac Stark shift of the molecular electronic terms is modeled by an
instantaneous uniform addition of momentum $\delta p$ to all molecules.
The kick is followed by formation of sharp ``tips'' on each branch
of the spiral (Fig. \ref{fig:Fig1}b). Due to the spread in their
relative velocities in the anharmonic potential the tips desynchronize
with time. Here, we use the ensemble-averaged time-dependent internuclear
separation, $\braket{R}$$\left(t\right)$ as an indicator for the
distribution spreading. The initial filamentation of the phase space
leads to apparent decaying oscillations of $\braket{R}$$\left(t\right)$
as the distribution spreads. The kick at $t=t_{k}$ gives rise to
instantaneous response in $\braket{R}\left(t\right)$ which then decays
as well, but at time $t\approx2t_{k}$ the tips meet again resulting
in a transient spatial inhomogeneity and an echo signal can be clearly
observed (see Fig. \ref{fig:Fig1}d). The described mechanism is similar
to the one responsible for formation of transverse beam echoes predicted
and observed in particle accelerators, see \cite{Stupakov1992,Spentzouris1996,Stupakov2013,Sen2018}.

In the case of the second mechanism, we assume that the kick depletes
the population in some spatial interval, but does not affect the vibrational
motion. In other words, the pulse carves out a strip from the phase
space distribution by instantaneously removing all the particles within
this strip. This process mimics optical absorption and creates a ``hole''
on each branch of the spiral. Figure \ref{fig:Fig1}c shows the phase
space distribution shortly after the instantaneous depletion within
the strip 5.5 a.u. \textless{} \textit{R} \textless{} 6.1 a.u.. These
holes move relative to each other and desynchronize with time (similar
to the tips discussed above). As in the previous case, the holes meet
again at $t=2t_{k}$, leading to the echo response. Figure \ref{fig:Fig1}e
shows the echo signal induced by the second mechanism. As noted, while
the single molecule wave packet cannot be described classically, the
classical phase-space analysis sheds light on the origin of the echo
and suggests the two mechanisms leading to its formation. Results
of a full quantum mechanical simulation are presented below.\\

\noindent \textbf{\textcolor{teal}{Experimental methodology.}} The
experiments were performed on $\mathrm{Ar}_{2}$ in an ultra-high
vacuum chamber of COLTRIMS \cite{Dorner2000,Ullrich2003}. In this
methodology, the kinetic energy release (KER) of the dissociated molecules
is extracted by analyzing the ions collected after ionization by a
strong ultrashort laser pulse (Fig. \ref{fig:Fig2}c). The $\mathrm{Ar}_{2}$
dimers are generated in a supersonic molecular beam by expanding room
temperature argon gas through a 30 $\mathrm{\mu m}$ nozzle under
a backing pressure of 1.6 bar. A fraction of about $1\%$ dimers was
produced with respect to the atomic monomers. The estimated vibrational
temperature is $T_{\mathrm{vib}}=10$ K, so it is safe to assume that
the entire population is in the ground state (see Methods section).
A femtosecond laser pulse, derived from a Ti:sapphire multi-pass amplifier
(790 nm, 25 fs, 10 kHz), is split and frequency doubled using a \textit{\textgreek{b}}-BBO
crystal to produce 395 nm pump and kick pulses, and a third pulse
at the fundamental wavelength of 790 nm to probe the vibrational dynamics
of the molecules. The peak intensities used in these experiments were
$2.0\times10^{14}\;\mathrm{W/cm^{2}},$ $1.1\times10^{13}\;\mathrm{W/cm^{2}}$
and $8.0\times10^{13}\;\mathrm{W/cm^{2}}$ for the pump, kick, and
probe pulses, respectively. The durations of the pump, kick, and probe
pulses (FWHM) were measured at the focus to be 75 fs, 110 fs, and
70 fs, respectively. The three pulses were collinearly recombined
and focused by a concave silver mirror ($f=75$ mm) onto the gas beam
inside the vacuum chamber. The time delays between the three pulses
were controlled by two motorized delay stages. The 3D momenta and
kinetic energy of the $\mathrm{Ar}^{+}$ produced in dissociation
of the $\mathrm{Ar}_{2}^{+}$ were reconstructed from the time-of-flights
(TOF) and positions of the impacts on the Multi-Channel Plate (MCP)
detector.

Based on the driving pressure, temperature of the gas source and the
detailed geometry of the jet system, the density of the dimers in
the laser beam focal volume is estimated to be $2\times10^{6}\;\mathrm{molecules/cm^{3}}$.
For the laser beam diameter at the focus of $\approx7$ $\mathrm{\mu m}$
with a Rayleigh length of $\approx8$0 $\mathrm{\mu m}$, there is
less than one molecule on average in our laser interaction volume
at any given time. With the low rate of events, there is much less
than a single event per laser shot. At the given laser rep-rate of
10 kHz, the molecules in the interaction volume are refreshed after
each laser shot. The COLTRIMS way of measurement, in particular the
low event rate and individual particle detection, guarantees that
only one molecule is measured in any given event. Furthermore, if
by chance, more than one molecule had been dissociated, due to the
low detection efficiency of the MCP, typically only one $\mathrm{Ar}^{+}$
fragment will be detected, leading, again, to a single molecule event.
Note that if (with very low probability in our experiments) two ionic
fragments were detected, the event is later\textbf{\emph{ }}rejected
in the off-line data analysis. The $\mathrm{Ar}^{+}$ ions coming
from the dissociation of the $\mathrm{Ar}_{2}^{+}$ have a kinetic
energy of a few eV and are clearly distinguishable from those coming
from ionization of the Ar monomers. The monomer events are excluded
by examining the kinetic energy of the detected $\mathrm{Ar}^{+}$.\\

The interaction of the laser pulses with $\mathrm{Ar}_{2}$ is depicted
in Fig. \ref{fig:Fig2}a. The ground electronic state of $\mathrm{Ar}_{2}^{+}$
is binding with a dissociation energy of 1.34 eV and equilibrium interatomic
separation $R_{0}^{+}=4.5$ a.u. \cite{Ar2PlusPotential}, which is
smaller than the equilibrium separation of $\mathrm{Ar}_{2}$ ($R_{0}=7.2$
a.u.) \cite{Ar2-ab-initio-gs}. The pump pulse at 395 nm creates a
vibrational wave packet in the $\mathrm{I}\left(1/2u\right)$ state
of $\mathrm{Ar}_{2}^{+}$ centered at the equilibrium distance $R_{0}$
(point A) of the neutral $\mathrm{Ar}_{2}$ dimer. To follow the vibrational
dynamics, the wave packet is then dissociated by a time delayed probe
pulse with a delay covering the range of -0.25 ps \textless{} $t_{p}$\textless{}
4 ps. The probe pulse (790 nm) ``catches'' the oscillating molecule
at point B, as illustrated in Fig. \ref{fig:Fig2}a, leading to dissociation
of $\mathrm{Ar}_{2}^{+}$ via $\mathrm{Ar}_{2}^{+}\rightarrow\mathrm{Ar}^{+}+\mathrm{Ar}$,
denoted as $\mathrm{Ar}_{2}\left(1,0\right)$ channel. The produced
$\mathrm{Ar}^{+}$ fragment has KER in the range of $0.7\;\mathrm{eV}\leq\mathrm{KER}\leq1.6\;\mathrm{eV}$.

\begin{figure}[h]
\begin{centering}
\includegraphics[width=0.9\columnwidth]{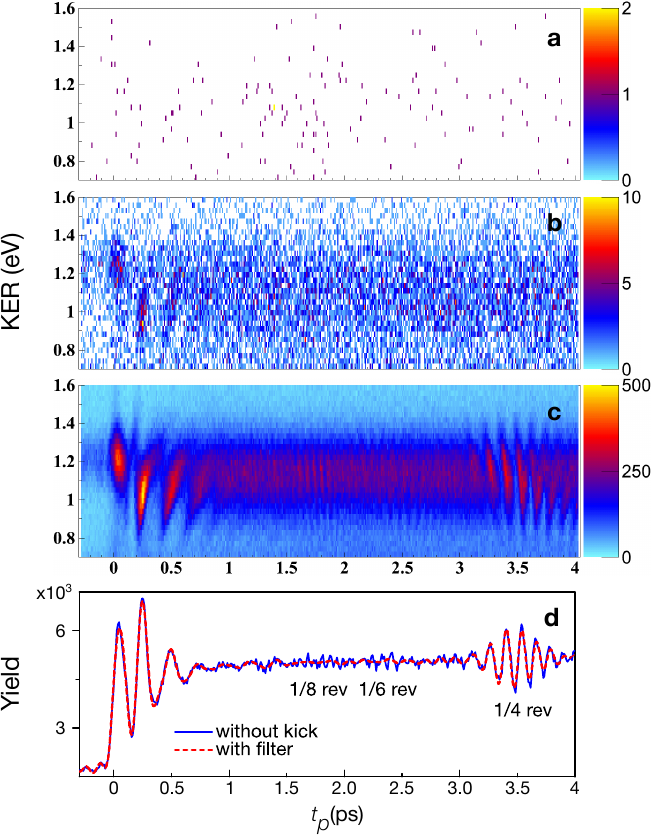}
\par\end{centering}
\caption{\textbf{KER distribution buildup. }Kinetic energy distribution of
molecular fragments ($\mathrm{Ar}_{2}\left(1,0\right)$ channel) as
a function of the probe delay $t_{p}$ following excitation by a single
femtosecond pulse at 790 nm. In panels (\textbf{a,b}) the events are
added: \textbf{a,} $150$ events, \textbf{b,} $1.5\times10^{4}$ events.
Panel\textbf{ c} depicts the full measurement after adding the single
molecule events from all 30 cycles ($1.5\times10^{6}$ events). The
instantaneous response to the pulse excitation is clearly visible,
as is the quarter revival around 3.5 ps. \textbf{d,} Yield of $\mathrm{Ar}_{2}\left(1,0\right)$
channel at energies ($0.7\;\mathrm{eV}\protect\leq\mathrm{KER}\protect\leq1.6\;\mathrm{eV}$)
induced by the probe pulse. Dashed red curve is the same data after
applying the low-pass filter (see text). \label{fig:Fig3}}
\end{figure}

In order to obtain a good signal-to-noise ratio in the measurement
of KER vs $t_{p}$, the measurements are repeated many times for different
time delays. Thus, the raw data consists of many uncorrelated single
events, and for each event KER and $t_{p}$ are recorded. In our experiments,
typically the time delay is scanned with 10 fs resolution. At each
time delay many laser shots are counted, and the entire sequence is
repeated till the desired signal-to-noise ratio is obtained. Figure
\ref{fig:Fig3}c is a result of 30 cycles scanning the whole range
of $t_{p}$, where each cycle consisted of $\approx8$ million laser
shots resulting in $\approx50,000$ usable events where one molecule
was dissociated and recorded at a time.

Figures \ref{fig:Fig3}(a-c) depicts the gradual buildup of the interference
pattern in the KER distribution as more and more events are accumulated.
Animated movies of this buildup may be seen in the Supplementary Movie.
The developing interference structure is a time-domain analog of the
buildup of spatial interference fringes while scattering single electrons
through a double slit (see \cite{Bach2013} and movies therein).

\begin{figure}
\begin{centering}
\includegraphics[width=0.9\columnwidth]{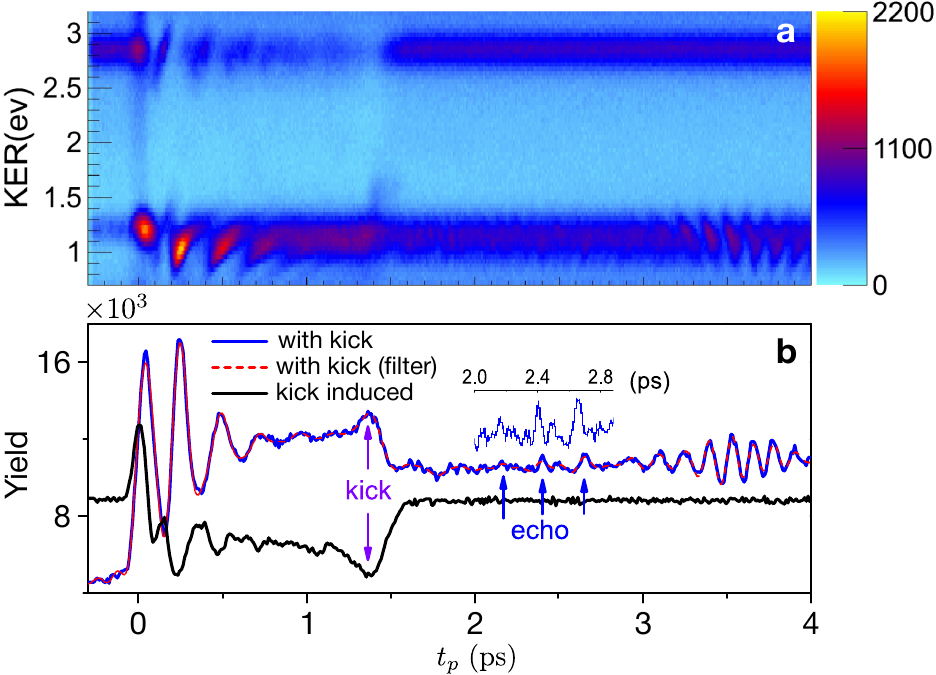}
\par\end{centering}
\caption{\textbf{KER distribution and yield of $\mathrm{Ar}_{2}\left(1,0\right)$
channel as a function of the probe delay in a presence of ``kick''
pulse.} Panels \textbf{a} and \textbf{b} are similar to Figs. 3(c,d),
but in the presence of kick pulse. The kick at 395 nm produces a higher
energy ($2.5\;\mathrm{eV}\protect\leq\mathrm{KER}\protect\leq3.2\;\mathrm{eV}$)
yield in the $\mathrm{Ar}_{2}\left(1,0\right)$ channel. Solid black
curve is the kick-induced yield. \label{fig:Fig4}}
\end{figure}

As the pump pulse ionizes an argon dimer, it creates a vibrational
wave packet that oscillates under the $\mathrm{I}\left(1/2u\right)$
potential. This wave packet is a superposition of multiple eigenstates
of the $\mathrm{I}\left(1/2u\right)$ potential as determined by the
original equilibrium state of the argon neutral dimer and the laser
parameters. The spatial localization of the wave packet persists for
several oscillations. However, due to the anharmonicity of the potential,
eventually dephasing takes over, leading to the collapse of the wave
packet. Figure \ref{fig:Fig3}d shows the yield of $\mathrm{Ar}_{2}\left(1,0\right)$
channel as a function of $t_{p}$. The curve is obtained by integrating
the KER distribution in the range $0.7\;\mathrm{eV}\leq\mathrm{KER}\leq1.6\;\mathrm{eV}$.
The blue curve in Fig. \ref{fig:Fig4}b reflects the vibrational motion
of the wave packet. The three pronounced oscillations with a period
of $\approx245\pm10$ fs near $t_{p}=0$ correspond to the initial
oscillatory motion of the still localized wave packet on the $\mathrm{I}\left(1/2u\right)$
potential (Fig. \ref{fig:Fig2}a). With the vibrational revival time
of $\mathrm{Ar}_{2}^{+}$ being 14 ps, the oscillations at doubled
frequency near probe delay of $t_{p}=3.5$ ps correspond to quarter-revival
\cite{Averbukh1989}, \cite{Jian2013}. Higher fractional revivals
are observed, as well, including the 1/8th revival near $t_{p}=1.8$
ps and the 1/6th revival near $t_{p}=2.3$ ps. The oscillation period
of the 1/8th revival is $63\pm10$ fs and that of 1/6th revival is
$82\pm10$ fs.\\

\noindent \textbf{Observation of echoes in a single molecule.} The
echo is induced by a kick pulse arriving at $t_{k}=1.36$ ps after
the pump. The kick pulse (395 nm) also couples the $\mathrm{I}\left(1/2u\right)$
and $\mathrm{II}\left(1/2g\right)$ states (at point C, Fig. \ref{fig:Fig2}a),
leading to partial dissociation of the same parent ion with fragments
having KER in the range of $2.5\;\mathrm{eV}\leq\mathrm{KER}\leq3.2\;\mathrm{eV}$,
higher than the energy band at $0.7\;\mathrm{eV}\leq\mathrm{KER}\leq1.6\;\mathrm{eV}$
induced by the probe and discussed above (see Fig. \ref{fig:Fig4}a).
The disjoint KER energy ranges enable unambiguous assignment of each
dissociation event to either the probe or the kick pulse. Figure \ref{fig:Fig4}b
shows the kick-induced (solid black) and probe-induced (solid blue)
yields of $\mathrm{Ar}_{2}\left(1,0\right)$ channel as a function
of probe delay, $t_{p}$. The kick perturbs the collapsed wave packet
and also partially transfers the population from the $\mathrm{I}\left(1/2u\right)$
state at point C to the $\mathrm{II}\left(1/2g\right)$ state, thus
triggering the two mechanisms for echo formation discussed above.

In addition to the prominent fractional revivals, the blue curve in
Figure \ref{fig:Fig4}b shows three new peaks corresponding to the
echo signal (showed in the inset and indicated by arrows at 2.16 ps,
2.41 ps and 2.66 ps). The period of these new oscillations is $\approx245\pm10$
fs, which is approximately the period of the original oscillations.
Thus, the kick has induced a delayed partial rephasing of the original
vibrational wave packet. For delays $0<t_{p}<t_{k}$ the probe precedes
the kick, therefore the corresponding kick induced (black curve) and
probe induced (blue curve) induced yields of the $\mathrm{Ar}_{2}\left(1,0\right)$
channel are out of phase. The drop in the probe induced yield (blue
curve) around $t_{p}=t_{k}$ results from the partial depletion of
the wave packet by the kick pulse. For longer probe delays $t_{p}>t_{k}$
the kick (at a fixed delay of $t_{k}=1.36$ ps) precedes the probe
and thus the kick-induced yield (black) remains constant.

Although the echo signal partially temporally overlaps the 1/6th fractional
revival (around probe delay $t_{p}=2.3$ ps), it can be distinguished
from the fractional revival because of their different oscillation
freqiencies. While the oscillation period of the echo signal is the
same as that of the original excitation, the frequency of the 1/6th
revival oscillations is three times that of the fundamental frequency.
By applying a low-pass filter, it is possible to significantly suppress
the contribution of the 1/6th revival. Details of the filtering procedure
are presented in the Methods Section. As is shown in Fig. \ref{fig:Fig3}d
(dashed red curve), the low-pass filter removes the high-frequency
components of signal, while retaining the contributions of the fundamental
and second-harmonic frequencies, which highlights the visibility of
the echo signal as displayed in Fig. \ref{fig:Fig4}b (dashed red
curve). The observation of these echoes is the main experimental result
of the current work.\\

\noindent \textbf{\textcolor{teal}{Quantum mechanical analysis of
the WPE.}} We carried out quantum mechanical simulations of the observed
WPE using two versions of the two-state molecular model. In the first
(simplified) one, we allowed dipole coupling between the $\mathrm{I}\left(1/2u\right)$
and the $\mathrm{II}\left(1/2g\right)$ states of $\mathrm{Ar}_{2}^{+}$
only for the kick pulse, and calculated the expectation value of the
interatomic separation as a function of time, $\braket{R}\left(t\right)$.
These results are used for comparing the quantum WPE in a single molecule,
and classical echo in an ensemble of many molecules. In the second,
more sophisticated model, both the kick and the probe pulses are allowed
to couple the two molecular states. This model fits well the experimental
configuration as described in Section III. Here, the probe pulse produces
an outgoing wave packet representing the products of dissociation
which is then analyzed for constructing the KER spectra as a function
of the probe delay. In both versions of the quantum model, we assume
that the initial state is prepared by instantaneous ionization of
$\mathrm{Ar}_{2}$ by the pump pulse, i.e. the ground state of the
neutral dimer is projected onto the $\mathrm{I}\left(1/2u\right)$
potential without any change. The time evolution of the two-state
system is described by a pair of coupled Schrödinger equations \cite{Garraway1995,Magrakvelidze2013}
\begin{equation}
\begin{cases}
i\dot{\psi_{1}} & =\left[T+V_{1}(R)\right]\psi_{1}-E(t)\mu_{\parallel}\left(R\right)\psi_{2}\\
i\dot{\psi_{2}} & =\left[T+V_{2}(R)\right]\psi_{2}-E(t)\mu_{\parallel}\left(R\right)\psi_{1}
\end{cases},\label{eq:coupled-system}
\end{equation}
where $T$ is the kinetic energy operator and indices 1, 2 refer to
the $\mathrm{I}\left(1/2u\right)$ and $\mathrm{II}\left(1/2g\right)$
potentials ($V_{1,2}(R)$), respectively. The off-diagonal terms describe
the dipole coupling between the two states and are equal to the product
of the coupling field $E(t)$ and the coordinate dependent transition
dipole moment $\mu_{\parallel}\left(R\right)$. In the numerical calculation
we add absorbing potentials to the diagonal terms (not shown here,
see Methods Section) to avoid spurious reflections from the edge of
the spatial grid. The electric fields of the kick and probe pulses
are given by
\begin{equation}
E_{j}(t)=E_{0j}\cos\left(\omega_{j}t\right)\exp\left[-2\ln2\left(\frac{t-t_{j}}{\mathrm{FWHM}_{j}}\right)^{2}\right],\label{eq:probe-field}
\end{equation}

\noindent where $E_{0k}$, $E_{0p}$ are the amplitudes of the fields,
$\omega_{k}$, $\omega_{p}$ are the carrier frequencies, $t_{k}$,
$t_{p}$ are the delays of the kick and probe pulses, respectively,
and $\mathrm{FWHM}_{k,p}$ refers to the temporal intensity profiles
of the kick and probe. The coordinate-dependent transition dipole
moment of $\mathrm{Ar}_{2}^{+}$ is $\mu_{\parallel}\left(R\right)=R/2$
\cite{Dipole1996} (see Methods Section for details). The coupling
field for the first version of the quantum model is $E(t)=E_{k}$,
while for the second one it is $E(t)=E_{k}+E_{p}$.

In the quantum simulations, we model the kick by adding an off-diagonal
term to the Hamiltonian describing the time evolution of the two-state
system. Such coupling acts as a source for the two echo-inducing mechanisms.
Unlike the classical case, here it is not possible to separate the
contributions of the momentum shift and depletion. However, as is
shown in \cite{Banin1994}, under the conditions of our experiment
the two state model does include both contributions - the creation
of the ``dynamical hole'' (depletion mechanism) and the momentum
shift (ac Stark shift mechanism). Details of the models and calculations
are provided in the Methods Section.

\begin{figure}
\begin{centering}
\includegraphics[width=0.9\columnwidth]{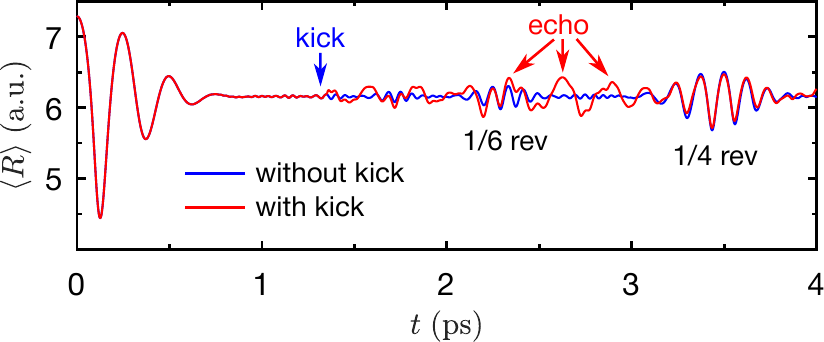}
\par\end{centering}
\caption{\textbf{Quantum mechanical simulations of the echo dynamics.} Expectation
value of the interatomic separation, $\braket{R}\left(t\right)\equiv\braket{\psi_{1}|R|\psi_{1}}/\braket{\psi_{1}|\psi_{1}}$
as a function of time with (blue) and without (red) the kick. This
result is calculated without the dissociating probe. The kick transfers
$\approx9\%$ of the population from the ground to the excited state.
The parameters of the kick pulse are: $I_{0k}=10^{12}\;\mathrm{W/cm^{2}}$,
$\omega_{k}=2\times0.0569$ a.u. (400 nm), $t_{k}=1.36$ ps and $\mathrm{FWHM}_{k}=10$
fs. \label{fig:Fig5}}
\end{figure}

\begin{figure}[h]
\centering{}\includegraphics[width=0.9\columnwidth]{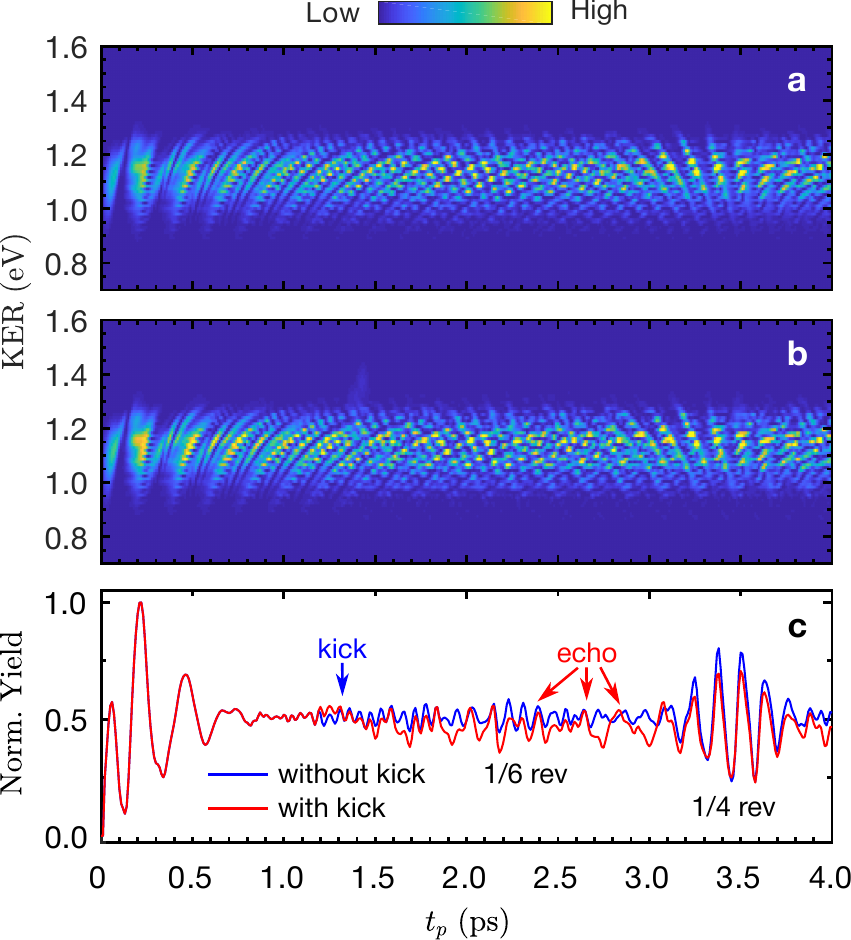}\caption{\textbf{KER distributions as a function of probe delay. }Kinetic energy
distribution of molecular fragments ($\mathrm{Ar}_{2}\left(1,0\right)$
channel) \textbf{a,} without and \textbf{b,} in the presence of the
kick pulse. The kick arrives at a delay $t_{k}=1.36$ ps. \textbf{c,}
The normalized probe-induced yield of the $\mathrm{Ar}_{2}\left(1,0\right)$
channel as a function of the probe arrival time, $t_{p}$, with (red
curve) and without (blue curve) the kick pulse. The parameters of
the probe pulse: $I_{0p}=10^{12}\;\mathrm{W/cm^{2}}$, $\omega_{p}=0.0569$
a.u. (800 nm) and $\mathrm{FWHM}_{p}=70$ fs. The kick pulse parameters
are the same as in Fig. \ref{fig:Fig5}. \label{fig:Fig6}}
\end{figure}

Figures \ref{fig:Fig5} and \ref{fig:Fig6} show the results of the
two simulations. Figures \ref{fig:Fig6}a and \ref{fig:Fig6}b display
the KER distributions as a function of the probe delay, $t_{p}$ without
(Fig. \ref{fig:Fig6}a) and with the kick pulse (Fig. \ref{fig:Fig6}b).
The normalized yield (Fig. \ref{fig:Fig6}c) is obtained by integrating
the KER distributions in the range of 0.7 eV \textless{} KER \textless{}
1.6 eV. In analogy to the classical analysis (Figs. \ref{fig:Fig1}(d,e)),
both Figs. \ref{fig:Fig5} and \ref{fig:Fig6}c show the change in
the periodicity of the signal around $t\approx2t_{k}$, corresponding
to the echo response.

In our simulations, the intensities of the pulses are smaller than
the experimentally measured ones. It is mainly because in the experiments
most of the molecules do not experience the maximal nominal intensity
of the pulse because of their location within the spatially Gaussian
intensity distribution. Also, we found that the 10 fs duration (broader
bandwidth) of the kick pulse results (for the adopted pulse intensities)
in a better visibility of the simulated echo signals (e.g. in Figs.
\ref{fig:Fig6}(b,c)), probably due to the fact that the pulses used
in the experiment also carry a broad bandwidth, but are longer due
to chirping. Further work will be required to address these issues
quantitatively.\textcolor{teal}{}\\

\noindent \textbf{\textcolor{teal}{Discussion.}} Here we introduce
WPE in a single isolated molecule and illustrate it in a vibrating
argon dimer cation. Unlike the conventional echoes, which are the
collective response of inhomogeneous ensembles of many molecules each
under different environmental conditions, here the entire echo cycle
occurs within a single isolated molecule. A coherent vibrational wave
packet is prepared by instantaneous ionization of a neutral argon
dimer, which then oscillates in the anharmonic ground state potential
of the cation. The spreading of the generated wave packet and the
disappearance of the initial coherent oscillations are due to the
dispersion of the vibrational frequencies within the excited wave
packet. After a time delay, a second impulsive excitation is applied,
inducing a subsequent echo response showing a (partial) recovery of
the initial coherent oscillations. This intramolecular dynamics was
measured by breaking the molecular ion in a COLTRIMS setup where the
kinetic energy of probe-induced dissociation products was measured
as a function of the probe delay. The essence of our detection method
is that only events where a single molecule (ion) has been exploded
are counted. The experimental results are compared with fully quantum
mechanical simulations, and analogy is drawn with a related classical
echo phenomenon. The classical phase space analysis reveals a connection
between the echo in highly excited vibrating molecules and transverse
echoes in accelerator beams of particles \cite{Stupakov1992,Spentzouris1996,Stupakov2013,Sen2018}.
The analysis suggests two mechanisms for the vibrational quantum WPE:
one is based on impulsive shaking of the molecular potential due to
non-resonant Stark effect, while the second one relies on impulsive
local depletion of the vibrational wave packet.\textcolor{red}{{} }

This classical analysis provides intuitive hints about the origin
of the vibrational echo, however a proper description of the WPE requires
a full quantum-mechanical treatment. It is interesting to compare
the present experiment with our previous studies on rotational alignment
echoes \cite{Lin2016}, which used a similar experimental approach.
In the latter case, the rotational temperature of the molecules was
high enough such that the dynamics could be treated classically, and
therefore the observed rotational alignment echo could be properly
described by the classical phase-space analysis. In the present case,
all the argon dimers are initially in the ground vibrational state,
and are then prepared in the same pure vibrational wave packet on
the ion potential. The only reason for the dephasing of this wave
packet, and its partial rephasing after a delayed kick is quantum
intra-molecular dynamics.

Recent years have shown increased interest in the coherent dynamics
of single quantum objects, especially in relation to the problem of
storing and retrieving quantum information in quantum network setups.
Studies on single electron spin echoes in molecular systems \cite{Orrit1995}
and isolated quantum dots \cite{Koppens2008,Yamamoto2010}, as well
as photon echoes from a few molecules \cite{Flemming2013}, were reported.
In these experiments, the environmental inhomogeneity is handled by
averaging the observations over many seconds during which the environment
changes many times, typically on a microsecond time scale. As is clearly
and explicitly pointed out by the authors \cite{Orrit1995,Flemming2013,Koppens2008,Yamamoto2010},
these ``single molecule'' time-averaged observations are completely
equivalent to standard echoes observed from an inhomogeneous ensemble
of many molecules, where the ensemble averages are replaced by time
averages over a single object. In contrast, in our experiments the
entire echo cycle of excitation and observation is completed in an
isolated single molecule on ultrafast time scale.

The present demonstration is done on simple dimer molecules, and the
method of detection is photon-induced dissociation followed by analysis
of the fragments impinging on an MCP detector. Naturally, such a detection
method is limited to a small number of particles hitting the detector,
and hence to small molecules. However, the phenomenon of WPEs in a
single molecule is generic, and with advanced detection methods, one
may envisage probing the internal dynamics of larger molecules, shedding
light on ultrafast intramolecular processes.

\section*{Methods}
\begin{widetext}
\noindent \textbf{Molecular beam.} Experimentally, a dilute molecular
beam was obtained by supersonicly expanding argon gas through a 30
$\mathrm{\mu m}$ nozzle into the ultrahigh vacuum chamber of the
COLTRIMS apparatus. As is common in similar molecular beam experiments,
the density of molecule in the focal volume is estimated by considering
the driving pressure and temperature of the gas source, the pumping
capability, and the geometry of the system. The vibrational temperature
of the molecules shares a similar value with translation temperature,
which can be estimated using $T_{\mathrm{vib}}\approx T_{\mathrm{trans}}=\text{\textgreek{D}}p^{2}/[4\ln(4)k_{\mathrm{B}}m]$
, where $k_{\mathrm{B}}$ is the Boltzmann constant, and $\Delta p$
and m are the FWHM of the momentum distribution in the jet direction
and mass of the singly ionized $\mathrm{Ar}_{2}^{+}$, respectively.
In our experiment, a momentum width in the jet direction was measured
to be $\Delta p\approx5.1$ a.u. of $\mathrm{Ar}_{2}^{+}$ cations
created by pump pulse polarized orthogonally to the gas jet. Thus,
the vibrational temperature in the interaction region of the argon
dimers is estimated to be $T_{\mathrm{vib}}=10$ K. At this temperature
it is safe to assume that the entire population is in the ground state.
\\

\noindent \textbf{Signal filtering.} We applied a low-pass filter
to curves in Figs. 3d and 4b. Using a discrete Fourier transform,
we obtained the frequency spectrum of the signal, containing the fundamental
vibration frequency, as well as quarter and 1/6th revivals frequencies,
denoted $f$, $2f$ and $3f$, respectively. Then we retain only the
frequencies below $2f$ in order to remove the 1/6th revival and higher
frequency components. Finally, we reconstructed the signal in the
time domain by applying the inverse discrete Fourier transform. Figure
\ref{fig:Fig7} shows the magnified portion of Figs. 3d and 4b.\\
\begin{figure}[H]
\centering{}\includegraphics[width=0.4\textwidth]{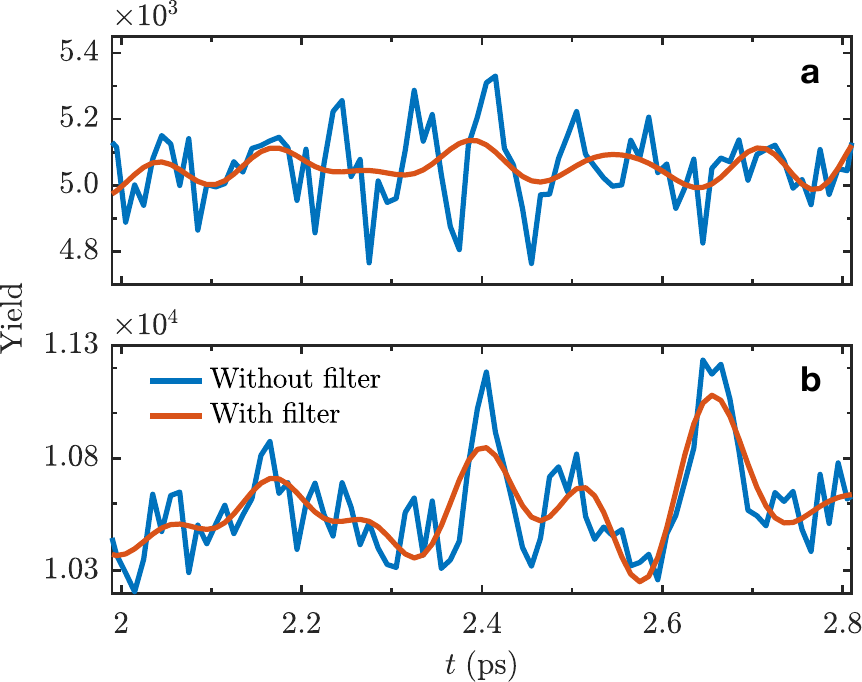}\caption{\textbf{Magnified portion of Figs. 3g and 4b. a,} Yield without the
kick pulse. \textbf{b,} Yield with the kick pulse. Both curves represent
the yield of $\mathrm{0.7\;eV\protect\leq KER\protect\leq1.6\;eV}$
channel. \label{fig:Fig7}}
\end{figure}

\noindent \textbf{Numerical scheme. }For solving the system of coupled
Schrödinger equations (Eq. 2), we discretize the spatial variable,
$R$ such that $\psi_{1}$ and $\psi_{2}$ are represented as column
vectors in $\mathbf{C}^{N}$ and $V_{1}$, $V_{2}$ are diagonal $N\times N$
matrices, where $N$ is the size of the spatial grid. The discrete
fourth order accurate approximation of the Laplacian operator is given
by
\begin{equation}
\frac{\partial^{2}\psi}{\partial R^{2}}\approx\frac{-\psi\left(R-2\Delta R\right)+16\psi\left(R-\Delta R\right)-30\psi\left(R\right)+16\psi\left(R+\Delta R\right)-\psi\left(R+2\Delta R\right)}{12\left(\Delta R\right)^{2}},\label{eq:Laplacian}
\end{equation}
and it is represented by a symmetric, sparse $N\times N$ matrix.
Here, $\Delta R$ is the spatial grid step size and we used $\Delta R=0.15$
a.u. The above form assumes that the wavefunction vanishes at the
boundaries, $\psi\left(R\leq R_{\mathrm{min}},t\right)=\psi\left(R\geq R_{max},t\right)=0$.
To assure this, we chose the grid size and the total propagation time
($t=t_{p}+0.3$ ps), such that the detached probe-induced wave packet
does not reach the boundary at all. We also included the absorbing
potential near the right edge of the spatial grid
\begin{equation}
V_{\mathrm{abs}}=-ia\Theta(R-R_{\mathrm{b}})\left(\frac{R-R_{\mathrm{b}}}{R_{\mathrm{max}}-R_{\mathrm{b}}}\right)^{2}.\label{eq:absorbing-potential}
\end{equation}
Here, $\Theta\left(\cdot\right)$ is a unit step function and we used
the following parameters: $a=0.01$ a.u., the right grid edge is $R_{\mathrm{max}}=34$
a.u., beginning of the absorbing layer lies at $R_{\mathrm{b}}=32$
a.u. The absorbing boundary is important when the probe pulse arrives
after the kick pulse and the kick-induced outgoing wave packet may
have enough time to reach the boundary. The off-diagonal elements
of the system (Eq. 2) are $E(t)\mu_{\parallel}\left(R\right)$. Strictly
speaking, the parallel transition dipole moment, $\mu_{\parallel}\left(R\right)$
equals to $R/2$ only for the potentials without the spin-orbit coupling
\cite{Dipole1996}. Nevertheless, we set $\mu_{\parallel}\left(R\right)=R/2$,
because near the resonance point B (Fig. \ref{fig:Fig2}a) $\mu_{\parallel}\approx R/2$.
After discretizing the right hand side of Eq. 2, ordinary differential
equation solver (based on the fourth order Runge-Kutta algorithm)
was used for time propagation.\\

\noindent \textbf{KER calculation.} To obtain the distribution of
the kinetic energy release (KER) as a function of the probe delay,
$\xi\left(\mathrm{KER},t_{p}\right)$ we allowed propagation of the
wave packet on both $\mathrm{I}\left(1/2u\right)$ and $\mathrm{II}\left(1/2g\right)$
potential curves up to $t_{\mathrm{max}}=t_{p}+0.3$ ps for each probe
pulse delay, $t_{p}$ and computed the momentum space representation
of the wavefunctions at this moment. The integration was done in the
region $R\in[10,R_{\mathrm{max}}]$, where the wave packets representing
the dissociation products are well separated from the bound part.
Then, the probability density for kinetic energy was obtained by
\begin{equation}
\xi\left(\mathrm{KER},t_{p}\right)\propto\frac{\sqrt{M}}{p}\left(\left|\tilde{\psi}_{\mathrm{I}\left(1/2u\right)}\left(p,t_{\mathrm{max}}\right)\right|^{2}+\left|\tilde{\psi}_{\mathrm{II}\left(1/2g\right)}\left(p,t_{\mathrm{max}}\right)\right|^{2}\right),\label{eq:KER-calculation}
\end{equation}
\end{widetext}

\noindent where the $\tilde{\cdot}$ denotes Fourier transform.

\section*{References}

\bibliographystyle{unsrt}

\section*{Acknowledgments}

We acknowledge useful discussions with Dan Oron, Dekel Raanan and
Gennady Stupakov. \textcolor{black}{This work is supported by the
National Key R\&D Program of China (Grant No. 2018YFA0306303), the
National Natural Science Foundation of China (Grant Nos. 11425416,
11834004, 61690224, 11621404 and 11761141004), the 111 Project of
China (Grant No. B12024), the Israel Science Foundation (Grant No.
746/15), the ICORE program ``Circle of Light'', and the ISF-NSFC
(Grant No. 2520/17).} I.A. acknowledges support as the Patricia Elman
Bildner Professorial Chair, and thanks for the hospitality extended
to him by the UBC Department of Physics\&Astronomy where part of this
work was carried out. This research was made possible in part by the
historic generosity of the Harold Perlman Family.
\end{document}